\documentclass[12pt] {article}
\usepackage{psfrag}
\usepackage{graphicx}
\usepackage{latexsym,amsfonts}
\usepackage{amssymb}
\begin{document}
\setcounter{page}{1}

\title{Bosonic vacuum wave functions from the BCS--type wave function
of the ground state of the massless Thirring model}

\author{M. Faber\thanks{E--mail: faber@kph.tuwien.ac.at, Tel.:
+43--1--58801--14261, Fax: +43--1--58801--14299} ~~and~
A. N. Ivanov\thanks{E--mail: ivanov@kph.tuwien.ac.at, Tel.:
+43--1--58801--14261, Fax: +43--1--58801--14299}~\thanks{Permanent
Address: State Polytechnical University, Department of Nuclear
Physics, 195251 St. Petersburg, Russian Federation}}

\date{\today}

\maketitle

\vspace{-0.5in}
\begin{center}
{\it Atominstitut der \"Osterreichischen Universit\"aten,
Arbeitsbereich Kernphysik und Nukleare Astrophysik, Technische
Universit\"at Wien, \\ Wiedner Hauptstr. 8-10, A-1040 Wien,
\"Osterreich }
\end{center}

\begin{center}
\begin{abstract}
A BCS--type wave function describes the ground state of the massless
Thirring model in the chirally broken phase. The massless Thirring
model with fermion fields quantized in the chirally broken phase
bosonizes to the quantum field theory of the free massless
(pseudo)scalar field (Eur. Phys. J. C {\bf 20}, 723 (2001)). The wave
functions of the ground state of the free massless (pseudo)scalar
field are obtained from the BCS--type wave function by averaging over
quantum fluctuations of the Thirring fermion fields. We show that
these wave functions are orthogonal, normalized and non--invariant
under shifts of the massless (pseudo)scalar field. This testifies the
spontaneous breaking of the field--shift symmetry in the quantum field
theory of a free massless (pseudo)scalar field. We show that the
vacuum--to--vacuum transition amplitude calculated for the bosonized
BCS--type wave functions coincides with the generating functional of
Green functions defined only by the contribution of vibrational modes
(Eur. Phys. J. C {\bf 24}, 653 (2002)) . This confirms the assumption
that the bosonized BCS--type wave function is defined by the
collective zero--mode (hep--th/0212226). We argue that the obtained
result is not a counterexample to the Mermin--Wagner--Hohenberg and
Coleman theorems. 
\end{abstract}
\end{center}

\newpage

\hspace{0.2in} A recent analysis of the massless Thirring model
\cite{FI1} has shown that the wave function $|\Omega\rangle$ of the
ground state of the massless Thirring model is of BCS form
\begin{eqnarray}\label{label1}
|\Omega\rangle = \prod_{\textstyle k^1}[u_{\textstyle k^1} +
 v_{\textstyle k^1}\, a^{\dagger}(k^1)b^{\dagger}(-k^1)]|0\rangle,
\end{eqnarray}
where $|0\rangle$ is the perturbative, chiral symmetric vacuum,
$a^{\dagger}(k^1)$ and $b^{\dagger}(-k^1)$ are creation operators of
fermions and anti--fermions with momentum $k^1$ and $-k^1$,
respectively, $u_{\textstyle k^1}$ and $ v_{\textstyle k^1}$ are
coefficient functions  \cite{FI1}
\begin{eqnarray}\label{label2}
u_{\textstyle k^1} = \sqrt{\frac{1}{2}\Bigg( 1 +
\frac{|k^1|}{\sqrt{(k^1)^2 + M^2}}\Bigg)}\quad,\quad v_{\textstyle
k^1} = \varepsilon(k^1)\,\sqrt{\frac{1}{2}\Bigg( 1 -
\frac{|k^1|}{\sqrt{(k^1)^2 + M^2}}\Bigg)},
\end{eqnarray}
where $\varepsilon(k^1)$ is the sign function.

The wave function (\ref{label1}) describes the fermions in a finite
volume $L$. According to Yoshida \cite{KY81} in the limit $L\to
\infty$ the wave function can be transcribed into the form
\begin{eqnarray}\label{label3}
|\Omega\rangle = \exp\Big\{
 \int^{\infty}_{-\infty}dk^1\,\tilde{\Phi}(k^1)\,
 [a^{\dagger}(k^1)b^{\dagger}(-k^1) - b(-k^1)a(k^1)]\Big\}\,|0\rangle,
\end{eqnarray}
where the phase $\Phi(k^1)$ is defined by
\begin{eqnarray}\label{label4}
 \tilde{\Phi}(k^1) = \frac{1}{2}\,\arctan\Big(\frac{M}{k^1}\Big).
\end{eqnarray}
Under chiral rotations
\begin{eqnarray}\label{label5}
&&\psi(x) \to {\psi\,}'(x) = e^{\textstyle
i\gamma^5\alpha_{\rm A}}\psi(x),\nonumber\\
&&\bar{\psi}(x) \to
{\bar{\psi}\,}'(x) = \bar{\psi}(x)\,e^{\textstyle
i\gamma^5\alpha_{\rm A}}, 
\end{eqnarray}
where $\psi(x)$ and $\bar{\psi}(x)$ are operators of the massless
Thirring fermion fields, the wave function (\ref{label3}) transforms
as follows \cite{FI1}
\begin{eqnarray}\label{label6}
\hspace{-0.5in}&&|\Omega\rangle \to |\Omega; \alpha_{\rm A}\rangle
=\nonumber\\ \hspace{-0.5in}&&=\exp\Big\{
\int^{\infty}_{-\infty}dk^1\tilde{\Phi}(k^1)[a^{\dagger}(k^1)
b^{\dagger}(-k^1)e^{\textstyle -2i\alpha_{\rm A}\varepsilon(k^1)} -
b(-k^1) a(k^1)e^{\textstyle +2i\alpha_{\rm A}\varepsilon(k^1)}]\Big\}
|0\rangle.
\end{eqnarray}
As has been shown in \cite{FI1} the wave functions $|\Omega;
\alpha_{\rm A}\rangle$ and $|\Omega; \alpha\,'_{\rm A}\rangle$
(\ref{label5}) are orthogonal for $\alpha\,'_{\rm A} \neq \alpha_{\rm
A}\,({\rm mod}\,2\pi)$, i.e. $\langle \alpha\,'_{\rm A};
\Omega|\Omega; \alpha_{\rm A}\rangle = \delta_{\alpha\,'_{\rm A}
\alpha_{\rm A}}$.

In the following we prove that the bosonized version of the wave
functions (\ref{label6}) are also orthogonal. We express
(\ref{label6}) in terms of the fermion field operators $\psi(x)$ and
$\bar{\psi}(x)$ using
\begin{eqnarray}\label{label7}
a(k^1) &=& \frac{1}{\sqrt{2\pi}}\int^{\infty}_{-\infty}dx^1\,
\frac{u^{\dagger}(k^1)}{\sqrt{2k^0}}\,\psi(0,x^1)\,e^{\textstyle
-ik^1x^1},\nonumber\\ b(-k^1)
&=&\frac{1}{\sqrt{2\pi}}\int^{\infty}_{-\infty}dy^1\,
\psi^{\dagger}(0,y^1)\,\frac{v(- k^1)}{\sqrt{2k^0}}\,e^{\textstyle
+ik^1y^1}.
\end{eqnarray}
The wave functions $u(k^1)$ and $v(k^1)$ are defined by \cite{FI1}
\begin{eqnarray}\label{label8}
u(k^1) = \sqrt{2k^0}{\displaystyle \left(\begin{array}{c}\theta(+k^1)
\\ \theta(-k^1)
\end{array}\right)}\quad,\quad v(k^1) = \sqrt{2k^0}{\displaystyle 
\left(\begin{array}{c} + \theta(+k^1) \\ - \theta(-k^1)
\end{array}\right)}.
\end{eqnarray}
They are normalized to $u^{\dagger}(k^1)u(k^1) =
v^{\dagger}(k^1)v(k^1) = 2k^0$ and $\theta(\pm k^1)$ are Heaviside
functions. Substituting (\ref{label7}) in (\ref{label6}) we get
\begin{eqnarray}\label{label9}
\hspace{-0.5in}|\Omega; \alpha_{\rm A}\rangle = \exp\Big\{
\int^{\infty}_{-\infty}dx^1\int^{\infty}_{-\infty}dy^1\,\Phi(x^1 -
y^1)\,\bar{\psi}(0,x^1)\gamma^5e^{\textstyle +2i\gamma^5\alpha_{\rm
A}}\psi(0,y^1)\Big\}|0\rangle,
\end{eqnarray}
where $\Phi(x^1 - y^1)$ is the original of $\tilde{\Phi}(k^1)$
\begin{eqnarray}\label{label10}
\hspace{-0.5in}\Phi(x^1 - y^1) &=& \int^{\infty}_{-\infty}
\frac{dk^1}{4\pi} \arctan\Big(\frac{M}{k^1}\Big)e^{\textstyle
+ik^1(x^1 - y^1)} = \nonumber\\
\hspace{-0.5in}&=& \int^{\infty}_{-\infty}
\frac{dk^1}{4\pi}\Big[\frac{\pi}{2} -
\arctan\Big(\frac{k^1}{M}\Big)\Big]\,e^{\textstyle +ik^1(x^1 - y^1)}=
\nonumber\\
\hspace{-0.5in}&=& \frac{\pi}{4}\,\delta(x^1 - y^1) + \int^{\infty}_0
\frac{dk^1}{2\pi i}\arctan\Big(\frac{k^1}{M}\Big)\,\sin(k^1(x^1 -
y^1)) =\nonumber\\
\hspace{-0.5in}&=&\frac{\pi}{4}\,\delta(x^1 - y^1) +
\frac{M}{x^1 - y^1}\int^{\infty}_0 \frac{dk^1}{2\pi
i}\,\frac{\cos(k^1(x^1 - y^1))}{(k^1)^2 + M^2}=\nonumber\\
\hspace{-0.5in}&=& \frac{\pi}{4}\,\delta(x^1 - y^1)+
\frac{1}{4i}\,\frac{\displaystyle e^{\textstyle -M|x^1 - y^1|}}{x^1 -
y^1} = \frac{1}{4i}\,\frac{\displaystyle e^{\textstyle -M|x^1 -
y^1|}}{x^1 - y^1 - i\,0}.
\end{eqnarray}
The function $\Phi(x^1 - y^1)$ has the property $\Phi^*(x^1 - y^1) =
\Phi(y^1 - x^1)$, which is important in order to provide a
normalization of the wave function (\ref{label9}), $\langle
\alpha_{\rm A}; \Omega |\Omega; \alpha_{\rm A}\rangle = 1$.

Integrating over $y^1$ \cite{FI1} we obtain
\begin{eqnarray}\label{label11}
\hspace{-0.5in}|\Omega; \alpha_{\rm A}\rangle =
\exp\Big\{\frac{\pi}{2}
\int^{\infty}_{-\infty}dx^1\bar{\psi}(0,x^1)\gamma^5e^{\textstyle
+2i\gamma^5\alpha_{\rm A}}\psi(0,x^1)\Big\}|0\rangle,
\end{eqnarray}
According to the Nambu--Jona--Lasinio condition \cite{NJL} the
operators of the massless fermion fields $\psi(0,x^1)$ at time zero
should be equal to the operators of the massive fermion fields
$\Psi(0,x^1)$ with dynamical mass $M$, i.e. $\psi(0,x^1) =
\Psi(0,x^1)$.  This yields
\begin{eqnarray}\label{label12}
\hspace{-0.5in}|\{\Psi\};\Omega; \alpha_{\rm A}\rangle &=& \exp\Big\{
\frac{\pi}{2}\int^{\infty}_{-\infty}dx^1\bar{\Psi}(0,x^1)
\gamma^5e^{\textstyle +2i\gamma^5\alpha_{\rm A}}\Psi(0,x^1)\Big\}
|0\rangle =\nonumber\\ &=& \exp\{i\,F[\Psi, \bar{\Psi}; \alpha_{\rm
A}]\} |0\rangle,
\end{eqnarray}
where the operator $\exp\{i\,F[\Psi, \bar{\Psi}; \alpha_{\rm
A}]\}$ is defined by
\begin{eqnarray}\label{label13}
\exp\{i\,F[\Psi, \bar{\Psi}; \alpha_{\rm A}]\} = \exp\Big\{
\frac{\pi}{2}\int^{\infty}_{-\infty}dx^1\bar{\Psi}(0,x^1)
\gamma^5e^{\textstyle +2i\gamma^5\alpha_{\rm A}}\Psi(0,x^1)\Big\}.
\end{eqnarray}
For convenience we have replaced $|\Omega; \alpha_{\rm A}\rangle$ by
$|\{\Psi\};\Omega; \alpha_{\rm A}\rangle$ in order to underscore that
the wave function of the ground state of the massless Thirring model
in the chirally broken phase is a functional of the dynamical fermion
fields.

As has been shown in \cite{FI1} the massless Thirring model bosonizes
to the quantum field theory of the free massless (pseudo)scalar field
$\vartheta(x)$. In order to find the wave function of the ground state
of the massless (pseudo)scalar field $\vartheta(x)$ we have to average
the operator (\ref{label13}) over the dynamical fermion degrees of
freedom. This can be carried out within the path--integral approach.
We would like to accentuate that the integration over fermion degrees
of freedom of the operator (\ref{label13}) should be understood as the
bosonization of the operator $\exp\{\,i\,F[\Psi,
\bar{\Psi};\alpha_{\rm A}]\}$ that means the replacement of fermion
degrees of freedom by boson ones, $\exp\{\,i\,F[\Psi,
\bar{\Psi};\alpha_{\rm A}]\} \to \exp\{\,i\,B[\vartheta; \alpha_{\rm
A}]\}$. The integration over fermion degrees of freedom does not touch
the wave function $|0\rangle$, which can be taken away in the
functional integral.

In the massless Thirring model the generating functional of Green
functions we define by \cite{FI1}
\begin{eqnarray}\label{label14}
\hspace{-0.3in}&&Z_{\rm Th}[J,\bar{J}] =\int {\cal D}\vartheta\,Z_{\rm
Th}[\vartheta; J,\bar{J}] = \nonumber\\
\hspace{-0.3in}&&= \int {\cal D}\vartheta\,{\cal D}\Psi\,{\cal
D}\bar{\Psi}\,\exp\Big\{i\int d^2x\,[\bar{\Psi}(x)(i\hat{\partial} -
M\,e^{\textstyle +i\gamma^5\vartheta(x)})\Psi(x) + \bar{J}(x)\Psi(x) +
\bar{\Psi}(x)J(x)]\Big\},\nonumber\\
\hspace{-0.3in}&& 
\end{eqnarray}
where $J(x)$ and $\bar{J}(x)$ are external sources of dynamical
fermions.

The bosonized version of the operator $\exp\{\,i\,F[\Psi,
\bar{\Psi};\alpha_{\rm A}]\}$ is defined by
\begin{eqnarray}\label{label15}
\hspace{-0.3in}&&\exp\{\,i\,B[\vartheta; \alpha_{\rm A}]\} =
\frac{1}{Z_{\rm Th}[\vartheta; 0,0]}\int {\cal D}\Psi {\cal
D}\bar{\Psi}\,\exp\{\,i\,F[\Psi, \bar{\Psi};\alpha_{\rm
A}]\}\nonumber\\
\hspace{-0.3in}&&\times\,\exp\Big\{i\int
d^2x\,[\bar{\Psi}(x)(i\hat{\partial} - M\,e^{\textstyle
+i\gamma^5\vartheta(x)})\Psi(x)+ \bar{J}(x)\Psi(x) +
\bar{\Psi}(x)J(x)\Big]\Big\}\Big|_{J = \bar{J} = 0}=\nonumber\\
\hspace{-0.3in}&& = \frac{1}{Z_{\rm Th}[\vartheta; 0,0]}\int {\cal
D}\Psi{\cal D}\bar{\Psi}\,\exp\{i\int
d^2x\,[\bar{\Psi}(x)\Big(i\hat{\partial} - M\,e^{\textstyle
+i\gamma^5\vartheta(x)}\,\Big)\Psi(x)\nonumber\\
\hspace{-0.3in}&& +
\frac{\pi}{2}\,\delta(x^0)\,\bar{\Psi}(x)\gamma^5\,e^{\textstyle
+2i\gamma^5\alpha_{\rm A}}\Psi(x)]\Big\},
\end{eqnarray}
By a chiral rotation $\Psi(x) \to e^{\textstyle
-i\gamma^5\vartheta(x)/2}\Psi(x)$ we obtain
\begin{eqnarray}\label{label16}
\hspace{-0.3in}&&\exp\{\,i\,B[\vartheta; \alpha_{\rm A}]\} =
\nonumber\\
\hspace{-0.3in}&& = \frac{1}{Z_{\rm Th}[\vartheta; 0,0]}\int {\cal
D}\Psi{\cal D}\bar{\Psi}\,\exp\Big\{i\int d^2x\,[\bar{\Psi}(x)
\Big(i\hat{\partial} + \frac{1}{2}\,\gamma^{\mu} \varepsilon_{\mu\nu}
\partial^{\nu}\vartheta(x) - M\Big)\Psi(x)\nonumber\\
\hspace{-0.3in}&& +
\frac{\pi}{2}\,\delta(x^0)\bar{\Psi}(x)\gamma^5\,e^{\textstyle
+i\gamma^5(2\alpha_{\rm A} - \vartheta(x))}\,\Psi(x)]\Big\}.
\end{eqnarray}
A possible chiral Jacobian, induced by this chiral rotation
\cite{FI1,FI2}, is canceled by the contribution of $Z_{\rm
Th}[\vartheta; 0,0]$ in the denominator.

Below we are going to show that integrating over fermion fields and
keeping leading terms in the large $M$ expansion \cite{FI1} we get the
following expression for the operator $\exp\{\,i\,B[\vartheta;
\alpha_{\rm A}]\}$
\begin{eqnarray}\label{label17}
\exp\{\,i\,B[\vartheta; \alpha_{\rm A}]\} =
\exp\Big\{i\,\frac{\pi}{2}\,\frac{M}{g}
\int^{\infty}_{-\infty}dx^1\,\sin\Big(\beta\vartheta(0,x^1) -
2\alpha_{\rm A}\Big)\Big\},
\end{eqnarray}
where we have also used the gap--equation for the dynamical mass $M$
(see Eq.(1.14) of Ref.\cite{FI1}). The coupling constant $\beta$ is
related to the coupling constant $g$ of the massless Thirring model
\cite{FI1,FI3}
\begin{eqnarray}\label{label18}
\frac{8\pi}{\beta^2} = 1 - e^{\textstyle - 2\pi/g}.
\end{eqnarray}
Expression (\ref{label17}) has been obtained as follows.  Integration
over the fermion fields gives the expression \cite{FI1}
\begin{eqnarray}\label{label19}
\hspace{-0.3in}&& \exp\{\,i\,B[\vartheta; \alpha_{\rm A}]\} =
\exp\Big\{i\int
d^2x\sum^{\infty}_{n=1}\frac{(-1)^{n-1}}{n}\frac{1}{4\pi}\int
\prod^{n-1}_{\ell}\frac{d^2x_{\ell}d^2k_{\ell}}{(2\pi)^2}\,e^{\textstyle
-ik_{\ell}\cdot(x_{\ell} - x)} \nonumber\\
\hspace{-0.3in}&&\times\int\frac{d^2k}{\pi i}\,{\rm
tr}\Big\{\frac{1}{M - \hat{k}}\Phi(x_1)\frac{1}{M - \hat{k} -
\hat{k}_1}\Phi(x_2)\ldots\Phi(x_{n-1})\frac{1}{M - \hat{k} -
\hat{k}_1 - \ldots - \hat{k}_{n-1}}\Phi(x)\Big\},\nonumber\\
\hspace{-0.3in}&&
\end{eqnarray}
where we have denoted
\begin{eqnarray}\label{label20}
\Phi(x_k) = -\frac{\pi}{2}\,\delta(x^0_k)i\gamma^5\, e^{\textstyle
i\gamma^5(2\alpha_{\rm A} - \vartheta(x_k))}
\end{eqnarray}
with $x_0 = x$ and $k = 0,1,2,\ldots, n-1$. For the subsequent
calculation of the momentum and space--time integrals we suggest to
use some kind of Pauli--Villars regularization. We smear the
$\delta$--functions $\delta(x^0_k)$ with the scales $M_j$
\begin{eqnarray}\label{label21}
\Phi(x_k; M_j) = -\frac{\pi}{2}\,f(M_jx^0_k)i\gamma^5\, e^{\textstyle
i\gamma^5(2\alpha_{\rm A} - \vartheta(x_k))}.
\end{eqnarray}
We require that in the limit $M_j \to \infty$ the function
$f(M_jx^0_k)$ converges to the $\delta$--function
$\delta(x^0_k)$. Introducing then the coefficients $C_j$, which
satisfy the constraints
\begin{eqnarray}\label{label22}
\sum^{N}_{j = 1} C_j = 1,\quad \sum^{N}_{j = 1} C_j M^n_j = 0\,,\, n =
1, 2, \ldots\,,
\end{eqnarray}
we rewrite (\ref{label19}) as follows
\begin{eqnarray}\label{label23}
\hspace{-0.3in}&&\exp\{\,i\,B[\vartheta; \alpha_{\rm A}]\}=
\exp\Big\{i\sum^N_{j = 1}C_j\int
d^2x\sum^{\infty}_{n=1}\frac{(-1)^{n-1}}{n}\frac{1}{4\pi}\int
\prod^{n-1}_{\ell}\frac{d^2x_{\ell}d^2k_{\ell}}{(2\pi)^2}\,e^{\textstyle
-ik_{\ell}\cdot(x_{\ell} - x)} \nonumber\\
\hspace{-0.3in}&&\times\int\frac{d^2k}{\pi i}\,{\rm
tr}\Big\{\frac{1}{M - \hat{k}}\Phi(x_1; M_j)\frac{1}{M - \hat{k} -
\hat{k}_1}\Phi(x_2; M_j)\ldots\Phi(x_{n-1}; M_j) \nonumber\\
\hspace{-0.3in}&&\times\frac{1}{M - \hat{k} - \hat{k}_1 - \ldots -
\hat{k}_{n-1}}\Phi(x; M_j)\Big\}.
\end{eqnarray}
Taking, first, the large $M$ expansion for the calculation of momentum
integrals \cite{FI1} and using the constraints (\ref{label22}) we get
(\ref{label17}).

Expression (\ref{label17}) can be obtained directly from
(\ref{label13}) by using our bosonization rules (see Eq.(3.24) of
Ref.\cite{FI1})
\begin{eqnarray}\label{label24}
\bar{\Psi}(0,x^1)\gamma^5\Psi(0,x^1) &=&
+\,i\,\frac{M}{g}\,\sin(\beta\,\vartheta(0,x^1)),\nonumber\\
\bar{\Psi}(0,x^1)\Psi(0,x^1) &=&
-\,\frac{M}{g}\,\cos(\beta\,\vartheta(0,x^1)).
\end{eqnarray}
Substituting (\ref{label24}) in (\ref{label13}) we arrive at
(\ref{label17}).

Using the operator (\ref{label17}) we define the bosonic wave function
\begin{eqnarray}\label{label25}
|\{\vartheta\};\Omega; \alpha_{\rm A}\rangle &=&
\exp\{\,i\,B[\vartheta; \alpha_{\rm A}]\}|0\rangle =\nonumber\\ &=&
\exp\Big\{i\,\frac{\pi}{2}\,\frac{M}{g}
\int^{\infty}_{-\infty}dx^1\,\sin\Big(\beta\vartheta(0,x^1) -
2\alpha_{\rm A}\Big)\Big\}|0\rangle,
\end{eqnarray}
which is normalized to unity. Now we have to show that such wave
functions are orthogonal for $\alpha\,'_{\rm A} \neq \alpha_{\rm
A}\,({\rm mod}\,2\pi)$
\begin{eqnarray}\label{label26}
\hspace{-0.3in}&&\langle \alpha\,'_{\rm A}; \Omega;
\{\vartheta\}|\{\vartheta\};\Omega; \alpha_{\rm A}\rangle =\Big\langle
0\Big|\exp\Big\{+i\,\pi\,\frac{M}{g}\,\sin(\alpha\,'_{\rm A} -
\alpha_{\rm A}) \nonumber\\
\hspace{-0.3in}&&\times\int^{\infty}_{-\infty}\!\!
dx^1\cos(\beta\,\vartheta(0,x^1) - \alpha\,'_{\rm A} - \alpha_{\rm
A})\Big\}\Big|0\Big\rangle = \lim_{L\to
\infty}\exp\Big\{+i\,\pi\,\frac{LM}{g}\,\sin(\alpha\,'_{\rm A} -
\alpha_{\rm A})\Big\}\nonumber\\
\hspace{-0.3in}&&\times \Big\langle 0\Big|\exp\Big\{- 2\pi
i\,\frac{M}{g}\,\sin(\alpha\,'_{\rm A} - \alpha_{\rm A})
\int^{\infty}_{-\infty}\!\!dx^1\sin^2\Big(\frac{\beta}{2}\,
\vartheta(0,x^1) - \frac{\alpha\,'_{\rm A} + \alpha_{\rm
A}}{2}\Big)\Big\}\Big|0\Big\rangle =\nonumber\\
\hspace{-0.3in}&&= \delta_{\alpha\,'_{\rm A}\alpha_{\rm A}}\Big\langle
0\Big|\exp\Big\{- 2\pi i\,\frac{M}{g}\,\sin(\alpha\,'_{\rm A} -
\alpha_{\rm A})
\int^{\infty}_{-\infty}\!\!dx^1\sin^2\Big(\frac{\beta}{2}\,
\vartheta(0,x^1) - \frac{\alpha\,'_{\rm A} + \alpha_{\rm
A}}{2}\Big)\Big\}\Big|0\Big\rangle=\nonumber\\
\hspace{-0.3in}&&= \delta_{\alpha\,'_{\rm A}\alpha_{\rm A}}.
\end{eqnarray}
As has been shown in \cite{FI1} the massless Thirring model with
fermion fields quantized in the chirally broken phase bosonizes to the
free massless (pseudo)scalar field theory with the Lagrangian ${\cal
L}(x) =
\frac{1}{2}\,\partial_{\mu}\vartheta(x)\partial^{\mu}\vartheta(x)$,
which is invariant under shifts of the field $\vartheta(x) \to
\vartheta\,'(x) = \vartheta(x) + \alpha$ with $\alpha \in
\mathbb{R}^{\,1}$ \cite{FI4}.

The wave function (\ref{label25}) describes the ground state of the
free massless (pseudo)scalar field $\vartheta(x)$. Since this wave
function is not invariant under the field--shifts, the symmetry is
spontaneously broken. The quantitative characteristic of the
spontaneously broken phase in the quantum field theory of the free
massless (pseudo)scalar field $\vartheta(x)$ is a non--vanishing
spontaneous magnetization ${\cal M} = 1$ \cite{FI4,FI5}. This confirms
fully the existence of the chirally broken phase in the massless
Thirring model obtained in \cite{FI1,FI2}--\cite{FI5}.

Let us consider the vacuum--to--vacuum transition amplitude for the
free massless (pseudo)scalar field theory with the ground state
described by the wave function (\ref{label25}) at $\alpha_{\rm A} = 0$
\begin{eqnarray}\label{label27}
|\{\vartheta\};\Omega \rangle =
\exp\Big\{i\,\frac{\pi}{2}\,\frac{M}{g}
\int^{\infty}_{-\infty}dx^1\,\sin\beta\vartheta(0,x^1)\Big\}|0\rangle,
\end{eqnarray}
The vacuum--to--vacuum transition amplitude we denote as $\langle
\Omega_+|\Omega_-\rangle_J$ \cite{JS70}, where $J(x)$ is an external
source of a free massless (pseudo)scalar field $\vartheta(x)$ and
$|\Omega_-\rangle$ and $\langle \Omega_+|$ are the vacuum states at
infinitely past $T = - \infty$ and at infinitely future $T = +
\infty$.

The time--dependent wave function (\ref{label27}) can be found in a
usual way using the translation formula \cite{CA68}
\begin{eqnarray}\label{label28}
&&|\{\vartheta\};\Omega_T \rangle = e^{\textstyle\,+ iH
T}|\{\vartheta\};\Omega \rangle = \nonumber\\ &&= e^{\textstyle\,+ iH
T}\exp\Big\{i\,\frac{\pi}{2}\,\frac{M}{g}
\int^{\infty}_{-\infty}dx^1\,\sin\beta\vartheta(0,x^1)\Big\}
e^{\textstyle\,-iH T}|0\rangle =\nonumber\\ &&=
\exp\Big\{i\,\frac{\pi}{2}\,\frac{M}{g}
\int^{\infty}_{-\infty}dx^1\,\sin\beta\vartheta(T,x^1)\Big\}|0\rangle,
\end{eqnarray}
where $H$ is the Hamilton function of the free massless (pseudo)scalar
field \cite{FI4,FI6}
\begin{eqnarray}\label{label29}
H = \frac{1}{2}\int^{\infty}_{-\infty}dx^1\,\Bigg[\Big(\frac{\partial
\vartheta(x)}{\partial x^0}\Big)^2 + \Big(\frac{\partial
\vartheta(x)}{\partial x^1}\Big)^2\Bigg].
\end{eqnarray}
Then, we have taken into account that $H|0\rangle = 0$ \cite{FI4,FI6}.

Denoting $|\Omega_-\rangle = |\{\vartheta\};\Omega_{-\infty} \rangle$
and $\langle \Omega_+| = \langle \Omega_{+\infty}; \{\vartheta\}|$ the
vacuum--to--vacuum transition amplitude $\langle
\Omega_+|\Omega_-\rangle_J$ for the wave function of the ground state
(\ref{label28}) is defined by
\begin{eqnarray}\label{label30}
&&\langle \Omega_+|\Omega_-\rangle_J = \int {\cal
D}\vartheta\,\exp\Big\{-\,i\,\frac{\pi}{2}\,\frac{M}{g}
\int^{\infty}_{-\infty}dx^1\,\sin(\beta\vartheta(+\infty,
x^1))\Big\}\nonumber\\
\hspace{-0.3in}&&\times\,\exp\Big\{i\int
d^2x\,\Big[\frac{1}{2}\partial_{\mu}\vartheta(x)
\partial^{\mu}\vartheta(x) +
\vartheta(x)J(x)\Big]\Big\}\nonumber\\
\hspace{-0.3in}&&\times\,\exp\Big\{+\,i\,\frac{\pi}{2}\,\frac{M}{g}
\int^{\infty}_{-\infty}dx^1\,\sin(\beta\vartheta(-\infty, x^1))\Big\}.
\end{eqnarray}
Recall, that the external source obeys the constraint
\begin{eqnarray}\label{label31}
\int d^2x\,J(x) = 0.
\end{eqnarray}
It allows to escape the excitation of the collective zero--mode. This
mode is responsible for the ``center of mass'' motion and originates
the infrared divergences in the quantum field theory of the free
massless (pseudo)scalar field \cite{FI4,FI5}.

The vacuum--to--vacuum transition amplitude (\ref{label30}) can be
transcribed into a more convenient form
\begin{eqnarray}\label{label32}
\langle \Omega_+|\Omega_-\rangle_J = \int {\cal
D}\vartheta\,\exp\Big\{i\int
d^2x\,\Big[\frac{1}{2}\partial_{\mu}\vartheta(x)
\partial^{\mu}\vartheta(x) + \vartheta(x)J(x) -
\lambda\,\frac{\partial}{\partial x^0}\sin(\beta\vartheta(x))
\Big]\Big\},
\end{eqnarray}
where $\lambda = \pi M/2 g$.

It is well--known that the total derivative with respect to time does
not contribute to the evolution of the system. Hence, the last term
can be dropped. This yields
\begin{eqnarray}\label{label33}
\langle \Omega_+|\Omega_-\rangle_J =
\exp\Big\{i\,\frac{1}{2}\int\!\!\!\int d^2x d^2y \,J(x)\,\Delta(x - y;
\bar{M})\,J(y)\Big\}.
\end{eqnarray}
The same result can be obtained by a direct calculation of the path
integral following \cite{FI2}.

The calculation of the path integral is very similar to that we have
carried our for the analysis of the renormalization of the
sine--Gordon model \cite{FI2} (see Appendix F of \cite{FI2}). We get
\begin{eqnarray}\label{label34}
\hspace{-0.3in}&&\langle \Omega_+|\Omega_-\rangle_J = \sum^{\infty}_{n
= 0}\frac{(-i \lambda)^n}{n!}\prod^n_{k = 1}\int
d^2x_k\,\frac{\partial}{\partial x^0_k}\int {\cal
D}\vartheta\,\prod^n_{k = 1}\sin\beta\vartheta(x_k)\nonumber\\
\hspace{-0.3in}&&\times \,\exp\Big\{i\int
d^2x\,\Big[\frac{1}{2}\partial_{\mu}\vartheta(x)
\partial^{\mu}\vartheta(x) + \vartheta(x)J(x)\Big]\Big\} =\nonumber\\
\hspace{-0.3in}&&= \exp\Big\{i\,\frac{1}{2}\int d^2x d^2y
\,J(x)\,\Delta(x - y; \bar{M})\,J(y)\Big\} +
\sum^{\infty}_{n=1}\frac{(-i
\lambda)^{2n}}{(2n)!}\,\frac{2n}{(2i)^{2n}}\prod^n_{k = 1}\prod^n_{j =
1}\nonumber\\ \hspace{-0.3in}&&\times\int\!\!\!\int d^2x_k d^2y_j
\,\frac{\partial}{\partial x^0_k}\frac{\partial}{\partial
y^0_j}\,\exp\Big\{i\,\beta^2\, n\,\Delta(0; \bar{M}) + i\,\beta^2\,
\sum^n_{j < k}\Delta(x_j - x_k; \bar{M}) + i\,\beta^2\nonumber\\
\hspace{-0.3in}&&\times\, \sum^n_{j < k}\Delta(y_j - y_k; \bar{M}) -
i\,\beta^2\, \sum^n_{k = 1}\sum^n_{j = 1}\Delta(x_k - y_j;
\bar{M})\Big]\Big\}\,\exp\Big\{i\,\beta\int d^2z\,\sum^n_{k =
1}(\Delta(x_k - z; \bar{M})\nonumber\\ \hspace{-0.3in}&& - \Delta(y_k
- z; \bar{M}))\,J(z) + i\,\frac{1}{2}\int\!\!\!\int
d^2z_1d^2z_2\,J(z_1)\,\Delta(z_1 - z_2; \bar{M})\,J(z_2)\Big\},
\end{eqnarray}
where $\bar{M}$ is a finite scale. One can easily show that the every
term of the infinite series vanishes due to the integration over
$x^0_k$ and $y^0_j$. There survives only the term, which does not
contain time derivatives.

The vacuum--to--vacuum transition amplitude (\ref{label33}) coincides
with generating functional of Green functions $Z[J]$ of the free
massless (pseudo)scalar field obtained in \cite{FI3}.
\begin{eqnarray}\label{label35}
Z[J] = \langle \Omega_+|\Omega_-\rangle_J =
\exp\Big\{i\,\frac{1}{2}\int\!\!\!\int d^2x d^2y \,J(x)\,\Delta(x - y;
\bar{M})\,J(y)\Big\}.
\end{eqnarray}
Due to the constraint (\ref{label31}) only the vibrational modes give
the contribution to the generating functional of Green functions
(\ref{label35}). This agrees with our recent assertion \cite{FI6} that
the ground state of the free massless (pseudo)scalar field is defined
by the collective zero--mode responsible for infrared divergences. The
collective zero--mode cannot be excited by the external source $J(x)$
satisfying the constraint (\ref{label31}). This make the quantum field
theory of a free massless (pseudo)scalar field infrared convergent.

As has been shown in \cite{FI4,FI5} the spontaneous magnetization,
calculated in the quantum field theory of a free massless
(pseudo)scalar field defined by the generating functional of Green
functions (\ref{label35}), does not vanish. It is equal to unity
${\cal M} = 1$ after renormalization of the ultra--violet divergences
\cite{FI4}.

We would like to emphasize that the obtained wave function of the
ground state of the bosonized version of the massless Thirring model
contradicts neither the Mermin--Wagner--Hohenberg theorem \cite{MWH}
nor Coleman's theorem \cite{SC73}. As has been pointed out by Mermin,
Wagner and Hohenberg \cite{MWH}, the vanishing of the {\it long--range
order}, i.e. the absence of the spontaneous breaking of continuous
symmetry, can be inferred only for non--zero temperature $T\neq 0$
\cite{MWH} and none conclusion about its value can be derived for
$T=0$ \cite{MWH}. Since spontaneous {\it magnetization} and fermion
{\it condensation}, caused by the BCS--type wave function
(\ref{label1}), have been found at $T = 0$, these results cannot be
considered as counterexamples to the MWH theorem. The same concerns
Coleman's theorem. Coleman's theorem asserts the absence of
spontaneous breaking of continuous symmetry in 1+1--dimensional
quantum field theories with Wightman's observables defined on the test
functions $h(x)$ from the Schwartz class ${\cal S}(\mathbb{R}^2)$. As
has been shown in \cite{FI4,FI5} in the quantum field theory of a free
massless (pseudo)scalar field $\vartheta(x)$, describing the bosonized
version of the massless Thirring model, Wightman's observables are
defined on the test functions $h(x)$ from the Schwartz class ${\cal
S}_0(\mathbb{R}^2) = \{h(x)\in {\cal S}(\mathbb{R}^2); \tilde{h}(0) =
0\}$, where $\tilde{h}(0)$ is the Fourier transform of $h(x)$ at
momentum zero. This agrees well with Wightman's assertion that quantum
field theory of a free massless (pseudo)scalar field acquires a
mathematical meaning only if Wightman's observables are defined on the
test functions from ${\cal S}_0(\mathbb{R}^2)$ \cite{AW64}. Thus, the
obtained results go beyond the scope of the applicability of Coleman's
theorem \cite{SC73} and cannot be considered as counterexamples to
this theorem \cite{FI4,FI5}.

\end{document}